\documentclass[11pt,twoside]{article}
\usepackage{macroaaa-eng} 
\usepackage{psfig} 
\usepackage{graphicx} 
 
\usepackage[T1]{fontenc} % Computer Modern (CM) fonts 
 
\usepackage{latexsym} 
\usepackage{verbatim} 
 
\begin{document} 
 
\vskip 1.0cm 
\markboth{A.~T.~Araudo et al.}{Non-thermal emission from massive YSOs} 
\pagestyle{myheadings}

\vspace*{0.5cm} 
\title{Non-thermal emission from massive YSOs. \\ 
Exploring the spectrum at high energies} 
 
\author{Anabella T. Araudo$^{1,2}$, Gustavo E. Romero$^{1,2}$, 
Valent\'i Bosch-Ramon$^{3}$ and
Josep M. Paredes$^{4}$} 
\affil{1 Instituto Argentino de 
Radioastronom\'{\i}a, (CCT La Plata - CONICET) 
Casilla de Corrreos No. 5, Villa Elisa 1894, Provincia de Buenos Aires
ARGENTINA.
E-mail: aaraudo@fcaglp.unlp.edu.ar
\\
~\\
2 Facultad de Cs. Astron\'omicas y Geof\'{\i}sicas, 
Universidad Nacional de La Plata, Paseo del Bosque, 1900 La Plata, 
Argentina 
\\
~\\
3 Max Planck Institut f\"ur Kernphysik, Saupfercheckweg
1, Heidelberg 69117, Germany
\\ 
~\\
4 Departament d'Astronomia i Meteorologia, Universitat de Barcelona, 
Mart\'{\i} i Franqu\`es 1, 08028, Barcelona, Spain \\
}

\begin{abstract}  
Thermal radio and X-ray emission has been traditionally associated with the 
formation of stars.  
However, in recent years,  non-thermal radiation from massive star  
forming regions has been detected.  
 
Synchrotron radio emission and non-thermal X-rays 
from the outflows of massive young stellar objects (YSOs) provide 
evidence of the presence of relativistic particles in these sources.  
 In YSOs, the acceleration of particles is likely produced where the thermal 
jet impacts on the surrounding medium and a shock wave is formed.  
Thus, particles might be accelerated up to relativistic 
energies through a Fermi-I type mechanism. 
  
 Relativistic electrons and protons can interact with thermal  
particles and photons, producing then $\gamma$-rays. These energetic photons  
could be detected by the new generation of instruments,  making massive  
YSOs  a new population of $\gamma$-ray surces.  
 
In the present contribution we briefly describe some massive star forming 
regions from which non-thermal radio emission has been detected. 
 In addition, we present a general model for high-energy 
radiation from the massive YSOs embedded in these regions. We take into 
account both leptonic and hadronic interactions of particles  
accelerated at the termination points of the collimated outflows 
ejected from the protostar.  
 
\end{abstract}

\section{Introduction} 
 
The mechanism of formation of massive stars ($M > 8M_{\odot}$) remains  
one of the open questions in astrophysics.
Massive stars appear in massive star associations where cloud 
fragmentation seems to be common.
It is known that these stars 
originate inside giant and massive molecular clouds but the sequence of 
processes that take place during the formation of the star are mostly 
unknown. It has been suggested, for example, that the coalescence of 
various protostars in the same cloud can lead to the emergence of a 
massive star (Bonnell et al. 1998). 
Alternatively, a massive star could form by the collapse 
of the core of a molecular cloud, with associated episodes of mass 
accretion and ejection, as observed in low-mass stars (Shu et al. 1987). 
In such a case, the effects of jets propagating 
through the medium that surrounds the protostar should be detectable. 
 
Until now, the formation of stars has been mostly associated with    
thermal radio and X-ray emission. However, 
non-thermal radio emission has been detected in some massive star  
forming regions. This is a clear evidence that efficient particle  
acceleration is occurring there, which may have as well a radiative  
outcome at energies much higher than radio ones. 
 
In the present contribution, based on recent multiwavelength  
observations and reasonable physical assumptions,  
we show that massive protostars 
could produce a significant amount of radiation in the gamma-ray domain,  
because of the dense and rich medium in which they are formed. 
 
\section{Non-thermal radio sources} 
 
In recent years, synchrotron radiation has been observed from some  
regions where massive stars form.  This 
emission is associated with outflows emanating from a central protostar. 
In what follows, we briefly describe some of 
these non-thermal radio sources that could be potential emitters of 
gamma-rays.  
 
\subsection{IRAS~16547$-$4247} 
 
 The triple radio source associated with the protostar IRAS~16547$-$4247   
is one of the best candidates to produce 
gamma-rays. This system is located  within a very dense region (i.e.  
densities $n\approx 5\times 10^5$~cm$^{-3}$) of 
a giant molecular cloud located at a distance of 2.9~kpc. The luminosity 
of the IRAS source is $L = 6.2\times 10^4 
L_{\odot}$, possibly being the most luminous known YSO associated with  
collimated thermal jets.  
 
ATCA and VLA observations (Garay et al. 2003, Rodr\'iguez et al.
2005) have shown that  the southern lobe of  
this system, of size $\sim 10^{16}$~cm, has a clear non-thermal spectrum, 
with an index $\alpha \sim 0.6$ ($S_{\nu} \propto \nu^{-\alpha}$).  
The specific flux of this source is 2~mJy at 
14.9~GHz and the estimated magnetic field is  $B \sim 2\times10^{-3}$~G  
(Araudo et al. 2007). 
 
\subsection{Serpens} 
 
The Serpens molecular cloud is located at a distance of $\sim$ 300 pc.  
One of the 
two central dense cores of this cloud is a triple radio source, composed  
by a central protostar (IRAS~18273$-$0113) and two lobes. The northwest (NW)  
hot-spot is connected with the central source by a highly collimated  
thermal jet,  
whereas the southeast (SE) is separated and broken into two clumps. 
The luminosity of the source IRAS~18273$-$0113 is  
 $L \sim 300 L_{\odot}$ and the particle density at the center of the  
molecular cloud is $n_0 \sim 10^5$ cm$^{-3}$. 
 
The observed radio emission (Rodr\'iguez et al. 1989, Curiel et al. 1993)
detected from the central and NW sources has a spectral index  
$\alpha \approx -0.15$ and $\alpha \approx 0.05$, respectively.  
This emission, of a luminosity $\sim 2-3$~mJy, is produced via thermal  
Bremsstrahlung.  
However, the radiation produced in the SE lobe seems to be 
non-thermal ($\alpha = 0.3$), likely produced via synchrotron emission. 
The specific flux of this source is $2-5$~mJy. The equipartition magnetic  
field estimated in the SE lobe is $B_{\rm{equip}} \sim 10^{-3}$~G  
(Rodr\'iguez et al. 1989).  
 
\subsection{HH 80-81} 
 
The famous Herbig-Haro objects called HH~80$-$81 are the south component of 
a system of radio sources located in the Sagitarius 
region, at a distance of 1.7 kpc.   
The central source has been identified with the luminous  
($L = 1.7\times 10^4 L_{\odot}$) protostar IRAS~18162$-$2048.  HH~80 North  
is the northern counterpart of HH~80$-$81. 
The velocity of the jet has been estimated as $v \sim 700$~km~s$^{-1}$,  
allowing to derive a dynamical age for the system similar to 4000~yr. 
 
Radio observations carried out with the VLA instrument 
(Mart\'i et al. 1993) showed that the central source has a spectral index  
$\alpha \sim 0.1$, typical of free-free emission, whereas HH~80$-$81 and 
HH~80 North are likely non-thermal sources, with spectral index  
$\alpha \sim 0.3$. The specific flux measured at a  
frecuency of 5 GHz is 
$F_{\nu}\sim 1-2$~mJy and $F_{\nu}\approx 2.4$~mJy for the sources 
HH~80$-$81 and HH~80 North, respectively. At this frecuency, the angular 
size of the north and the south components are $\sim 6^{\prime\prime}$. 
 
In addition, the HH~80$-$81 system is a source of thermal X-rays with  
a luminosity $L_X \sim 4.3\times10^{31}$~erg~s$^{-1}$ (Pravdo et al. 2004). 
 
\subsection{W3(OH)} 
 
Another interesting source to study is the system composed by an H$_2$O maser 
complex  and the Turner-Welch (TW) source in the W3 region (Wilner et al. 
1999). The central source of this system is a very luminous  
($L \sim 10^5 L_{\odot}$) YSO and 
the mean density of cool particles is  $n \sim 4\times10^4$~cm$^{-3}$. The 
distance to W3(OH) is 2.2~kpc. 
 
Continuum radio observations (Wilner et al. 1999) show the presence of a  
sinuous double-sided 
jet, emanating from the TW source. The observed radio flux, from 1.6 to 
15 GHz, is in the range 2.5-0.75 mJy, and the spectral index of the 
observed emission is clearly non-thermal: $\alpha = 0.6$.  
The inhomogeneous synchrotron model proposed by Reid et al. (1995)
predicts for the emitting jet a density of relativistic electrons  
$n_e(\gamma, r) = 0.068 \gamma^{-2}(r/r_0)^{-1.6}$~cm$^{-3}$ and a  
magnetic field  $B(r) = 0.01 (r/r_0)^{-0.8}$~G,  
where $r_0 = 6.6\times10^{15}$~cm, and $r$ and $\gamma$ are the distance  
to the jet origin and the electron Lorentz factor, respectively. Unlike  
in the previous cases, here the non-thermal  
radio emission comes from the jet and not from its termination region.  
Non-thermal jets associated with a YSO are uncommon.   
  
\section{Acceleration of particles and losses} 
 
The non-thermal radio emission observed in some massive star forming 
regions is interpreted as synchrotron radiation produced by the 
interaction of relativistic electrons with the magnetic field present 
in the cloud, being typically $B_{\rm{cloud}} \sim 10^{-3}\;\rm{G}$  
(Crutcher 1999).  These particles 
could be accelerated at a shock, formed in the point where the jet 
terminates, via  diffusive shock acceleration (Drury 1983).   
The acceleration efficiency, characterized by $\eta$, is related to  
the velocity  of the shock. Using the values of the velocities 
given for the sources 
described in the previous section, and assuming Bohm diffusion,  
values for $\eta$ of $\sim 10^{-6}-10^{-5}$ are obtained. 
 
Particles accelerated up to relativistic energies interact with the  
different fields present in the medium. As 
noted at the beginning of this section, electrons radiate synchrotron  
emission under the  
ambient magnetic field $B$. In addition, particles, 
electrons and protons, can also interact with the cold matter in  
the jet termination region (via relativistic 
Bremsstrahlung the leptons, and inellastic proton-nuclei collisions 
the protons).  
In addition, electrons interact with the 
background field of IR photons of the protostar, of energy density  
$u_{\rm{ph}}$, through inverse Compton (IC) scattering. 
 
\begin{figure}   
\begin{center} 
\hspace{0.25cm} 
   \psfig{figure=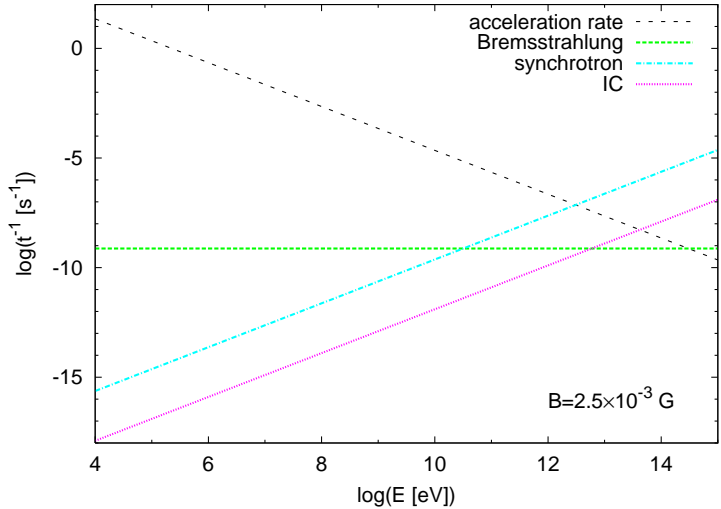,angle=270,width=10.cm} 
\caption{Energy loss and acceleration rates for electrons in the  
IRAS~16547$-$4247 southern lobe.} 
\label{fig1} 
\end{center} 
\end{figure}

Using the following parameter values:  
$n=5\times10^{5}$~cm$^{-3}$; $B=2.5-3\times10^{-3}$~G;  
and $u_{\rm ph}=3.2\times10^{-9}$~erg~cm$^{-3}$ given for  
IRAS~16547$-$4247 (Garay et al. 2003, Araudo et al. 2007), 
we  estimate the cooling time of the main leptonic processes in this  
scenario. As seen in Figure \ref{fig1}, 
relativistic Bremsstrahlung losses are dominant up to $\sim 10$~GeV.  
In addition, it is possible to see from 
this figure that the maximum energy achieved by electrons is  
$E_e^{\rm{max}} \sim 4$~TeV and is determined by 
synchrotron losses, being IC losses negligible.

Protons can be accelerated by the shock in the same way as electrons and  
interact with cold particles present in the 
cloud. The maximum energy achieved by protons is higher,  
$E_p^{\rm{max}} \sim 10^2$~TeV, and determined by the size of the 
acceleration region. In $pp$ interactions, besides 
$\gamma$-rays, secondary electron-positron pairs are produced.  
The maximun energies of these leptons is $E_{e^{\pm}}^{\rm{max}} \sim 10$~TeV.
These secondary particles will radiate by the same 
mechanisms as primary electrons (i.e.  synchrotron radiation, IC  
scattering and relativistic Bremsstrahlung). 
 
\section{High-energy emission} 
 
In order to calculate the non-thermal spectral energy distribution (SED) of 
a massive protostar, the magnetic  field of the cloud and the  
distributions of relativistic particles, $n(E)$, are needed.  
To obtain the magnetic field $B$ and the normalization 
constants of the particle distributions, we use the standard 
equations given by Ginzburg \& Syrovatskii (1964) for the observed  
synchrotron flux and assume equipatition between the magnetic  
and the relativistic particle energy densities: 
\begin{equation} 
\frac{B^2}{8\pi} = u_e + u_p + u_{e^{\pm}},  
\label{equip} 
\end{equation} 
where $u_e$ , $u_p$ and $u_{e^{\pm}}$ are the energy density of relativistic
electrons, protons and secondary pairs, respectively. 
In Eq. (\ref{equip}), the following relationships are implicit:  
$u_{\rm p}= a\,u_{\rm e}$ and  $u_{e^\pm}= f\,u_{\rm p}$. The constant
$a$ is a free  parameter of the model, that characterizes
how much energy goes to the different types of accelerated particles
and  $f$ can be  
estimated using the average ratio of the number of secondary pairs to  
$\pi^0$-decay photons as in Kelner et al. (2006). 

\begin{figure}   
\begin{center} 
\hspace{0.25cm} 
   \psfig{figure=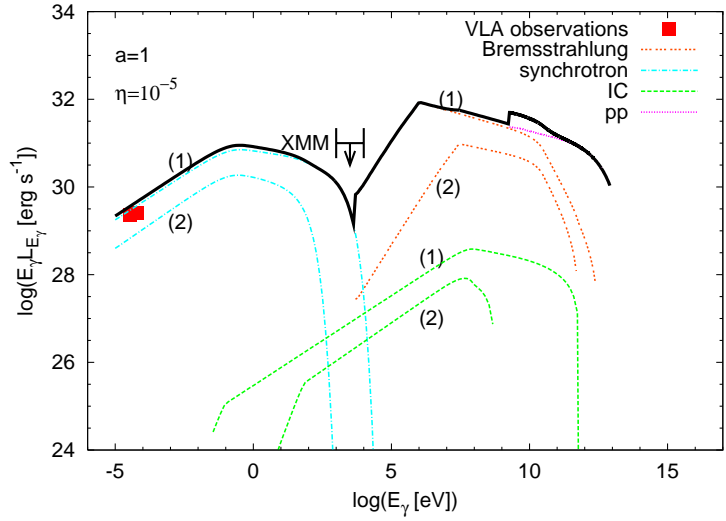,angle=270,width=10.cm} 
\caption{Spectral energy distribution for the south lobe of the 
YSO embedded in the source IRAS~16547$-$4247, for the case $a = 1$. 
The radiative contribution of 
secondary pairs $(2)$ produced via $\pi^\pm$-decay is shown along with  
the contribution of primary electrons $(1)$.} 
\label{fig2} 
\end{center} 
\end{figure}

\begin{figure}   
\begin{center} 
\hspace{0.25cm} 
   \psfig{figure=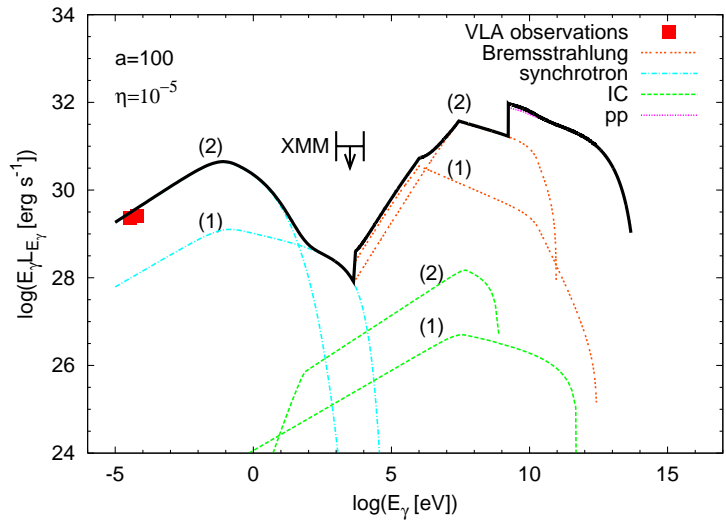,angle=270,width=10.cm} 
\caption{The same as in  Fig. \ref{fig2}, but for the case $a = 100$.} 
\label{fig3} 
\end{center} 
\end{figure}

In Figs. \ref{fig2} and \ref{fig3}, we show the computed broadband  
SEDs for the parameters of the source  IRAS~16547$-$4247, in the cases with
$a = 1$ (equipartition between protons and electrons) and $a = 100$
(larger acceleration efficiency in protons than in electrons). 
As seen from  these figures, in the former 
case the leptonic emission is dominated by the primary electron population
and in the latter by the secondary pairs. The typical lepton luminosities
are $L_{e} \sim 10^{32}$ erg s$^{-1}$.
Regarding the hadronic emission, 
the luminosity  produced by $\pi^0$-decay is higher in the case $a = 100$,
reaching a value of $L_{pp} \sim 5\times 10^{32}$ erg s$^{-1}$.  

\section{Discussion} 
 
In this work we show that, if the source is located at few kpc,  
the high-energy emission may be 
detected by GLAST and even by forthcoming Cherenkov telescope arrays after  
long enough exposure.  This opens a new 
window to the study of star formation and related processes. Also,  
determinations of the particle spectrum and its high-energy 
for different sources with a variety of environmental conditions  
can shed light on the properties of 
galactic supersonic outflows, and on the  particle acceleration  
processes occurring at their termination points.  
 
Radio observations already demonstrate that relativistic electrons are  
produced in some sources. According to the 
presence of non-thermal emission detected at cm-wavelengths and IR  
observations of the protostar emission we can 
suggest several good candidates to be targeted by GLAST. These objects  
are IRAS~16547$-$4247 (Araudo et al. 2007),  
the multiple  radio source in Serpens (Rodr\'iguez et al. 1989, 
Curiel et al. 1993), HH~80$-$81 (Mart\'i et al 1993) and 
W3 (Reid et al. 1995, Wilner et al. 1999). 

To conclude, we emphasize that massive YSO with bipolar outflows and  
non-thermal radio emission can form a new 
population of gamma-ray sources that could be unveiled by the next  
generation of $\gamma$-ray instruments. 
 
\acknowledgments  

A.T.A. and G.E.R. are supported by CONICET (PIP 5375) and the
Argentine agency ANPCyT through Grant PICT 03-13291 BID 1728/OC-AC.
V.B-R. gratefully acknowledges support from the Alexander von Humboldt
Foundation.
V.B-R., and J.M.P  acknowledge support by DGI
of MEC under grant AYA2004-07171-C02-01, as well as partial support by
the European Regional Development Fund (ERDF/FEDER).

\end{document}